# An Intelligent Material with Chemical Pathway Networks

Li Lin[1*] and Michael Keidar[1*]

[1] *Department of Mechanical and Aerospace Engineering, School of Engineering and Applied Science, The George Washington University*

*Corresponding Authors:* lilin@gwu.edu, keidar@gwu.edu

## Abstract

A new type of material with embedded intelligence, namely "intelligent plasma", is introduced. Such new material exhibits programmable chemical pathway networks resembling artificial neural networks. As a Markov process of chemistry, the chemical pathway network can be customized and thus the intelligent plasmas can be programmed to make their own decisions to react to the dynamic external and internal conditions. It finally can accomplish complex missions without any external controls from the humans while relying on its preprogrammed chemical network topology before the mission. To that end, only basic data input and readings are required without any external controls during the mission. The approach to "if" conditions and "while" loops of the programmable intelligent plasmas are also discussed with examples of applications including automatic workflows, and signal processing.

# Introduction

About seventy years after the age of Alan Turing, modern artificial intelligence (AI) is now able to defeat human champions of board games and complex video games where the AI can continuously decide on $10^{26}$ choices in real-time[1,2]. At the same time, AI algorithms stand on modern digital computer technologies limited by Moore's law[3-5]. Fortunately, it is possible to avoid such a dilemma. A material with an embedded chemical-based algorithm could be a solution. In the macroscopic view of massive molecular collisions, chemical reactions are Markov processes that a consecutive state of the system depends on the state in the previous moment only and is independent of the history. The concentration of each species in a chemical system is determined by the rates of inelastic collisions. In other words, the chemical compositions are determined by the set of rate equations which makes the mathematical relations among them similar to a feedforward or recursive computation of neuron values in an artificial neural network (ANN). The species concentrations and other parameters such as temperatures are the neuron values, and the rate coefficients are the weights among neurons. Therefore, it is attractive to program a multi-species chemical system based on the chemical pathway networks (CPN), and finally achieve an embedded intelligence to make an intelligent material that can react to the environment or process data without external manual controls. This makes a material an analog computer which may significantly impact the industry and scientific research communities, considering the advantages of recently revived modern analog computers comparing with the digital ones[6-8].

Recently, some perspective papers indicated future materials that can react to the environment automatically including a swarm of massive particles with self-organization capabilities, soft materials with embedded memories, adaptive abilities, and even microsystems[9,10]. These examples reveal the potential of intelligent materials for life science and bio-cybernetics[9]. However, there is another promising intelligent material with a border view of applications missing in these perspectives. Following both the magnetohydrodynamics and chemistry, plasmas could be ideal media to carry AI due to their high level of adaptation with more flexible parameters to be programmed. Modern plasma technologies also provide significant diverse applications including aerospace, material processing, metamaterials, biomedicine, communication, and sensors[11-20]. Therefore, once plasma systems can be designed and become an intelligent material, namely "intelligent plasmas", massive modern industrial and scientific applications would be benefitted. In the following sections, we propose a practical approach based on a theory of programmable chemical pathway networks to achieve such a new intelligent material.

## Theory

**Programmable Chemical Pathway Network.** The intelligence of materials means the ability to react properly after sensing the environmental conditions[9]. This includes data processing that outputs a signal based on the input signal, and automatically optimal controls according to the environmental perturbations. In a chemical system, temporally resolved species compositions are determined by the rate equations

$$\frac{\partial n_\zeta}{\partial t} = \sum_\eta k_\eta \prod_\xi n_{\eta,\xi}^\Delta \tag{1}$$

where $n_\zeta$ is the $\zeta^{th}$ species concentration to solve, $k_\eta$ is the rate coefficient of $\eta^{th}$ chemical reaction, and $n_{\eta,\xi}$ is the $\xi^{th}$ reactant density of the reaction with its reaction order $\Delta$. Usually, Eq. (1) can be solved using a matrix method for multiple-species systems[21]. Relying on the matrix multiplications, Eq. (1) is equivalent to

$$\frac{\partial \boldsymbol{N}_{s\times 1}}{\partial t} = (\boldsymbol{\Phi}_{s\times r} - \boldsymbol{\Gamma}_{s\times r})\boldsymbol{K}_{r\times 1} = \mathcal{F}(\boldsymbol{T}_{s\times 1}, \boldsymbol{N}_{s\times 1}) \tag{2}$$

where the matrix $\boldsymbol{N}$ is $s$ (species number) by 1 as a set of $n_\zeta$, the matrices $\boldsymbol{\Phi}$ and $\boldsymbol{\Gamma}$ are both $s$ by $r$ (chemical reaction number), and the matrix $\boldsymbol{K}$ is $r$ by 1, as the contribution list of each chemical reactions. Each contribution value is a product of the rate coefficient of the reaction with all the concentrations of reactants. The matrix $\boldsymbol{\Phi}$ represents the number of product molecules (columns) in each reaction (rows), as same as the matrix $\boldsymbol{\Gamma}$ which represents the reactants. Therefore, the term $\boldsymbol{\Phi} - \boldsymbol{\Gamma}$ equals the net increment of each species in each reaction with zeros for the species not involved. The matrix operation can be summarized as a mathematical operator $\mathcal{F}$ mapping between the input and output functional space, taking the species temperature set $\boldsymbol{T}$ and the concentration set $\boldsymbol{N}$ as inputs. Eq. (2) thus leads to the status of the system before and after a short period d$t$:

$$\boldsymbol{N}_{output} = \boldsymbol{N}_{input} + \mathcal{F}(\boldsymbol{T}_{input}, \boldsymbol{N}_{input})\mathrm{d}t \qquad (3)$$

Therefore, once the operator $\mathcal{F}$ can be designed properly to make the system able to assess the condition and output correctly, the system acquires intelligence according to the aforementioned definition. The $\boldsymbol{\Phi} - \boldsymbol{\Gamma}$ term indicates the topology of CPN where the connections can be customized in $\boldsymbol{K}$.

Each element in the matrix $\boldsymbol{K}$ contains two parts: the rate coefficient and the product of all reactant concentrations of each reaction. In the view of ANN, there is an activation function hidden in $\boldsymbol{K}$. For example, a reaction rate coefficient can obey the Arrhenius equation

$$k = A \exp\left(-\frac{E_a}{T}\right) \qquad (4)$$

where $A$ is a coefficient, $E_a$ is the activation energy required for the chemical reaction, and $T$ is the mean temperature of the reactants. As shown in Fig. 1a, $E_a$ plays as a threshold to determine if $T$ is high enough to results in a high $k$ value (activated).

Such an activation relationship is similar to the collector-emitter current *Ic* versus the collector-emitter voltage *Vc* controlled by the gate-emitter voltage *Vg* of an electric switch transistor such as an insulate-gate bipolar transistor (IGBT) as shown in Fig. 1b. Considering that a rate coefficient is a switch to control the reactants producing the products in a particular reaction, the mathematical role of it in a CPN is the same as a transistor in an integrated circuit such as a microprocessor unit (MPU). There is no doubt that logic gates exist in a substantially connected CPN, similar to the semiconductor components. Therefore, it is not surprising that a CPN can work as a programmable processor containing logic gates, and the basis to carry an algorithm.

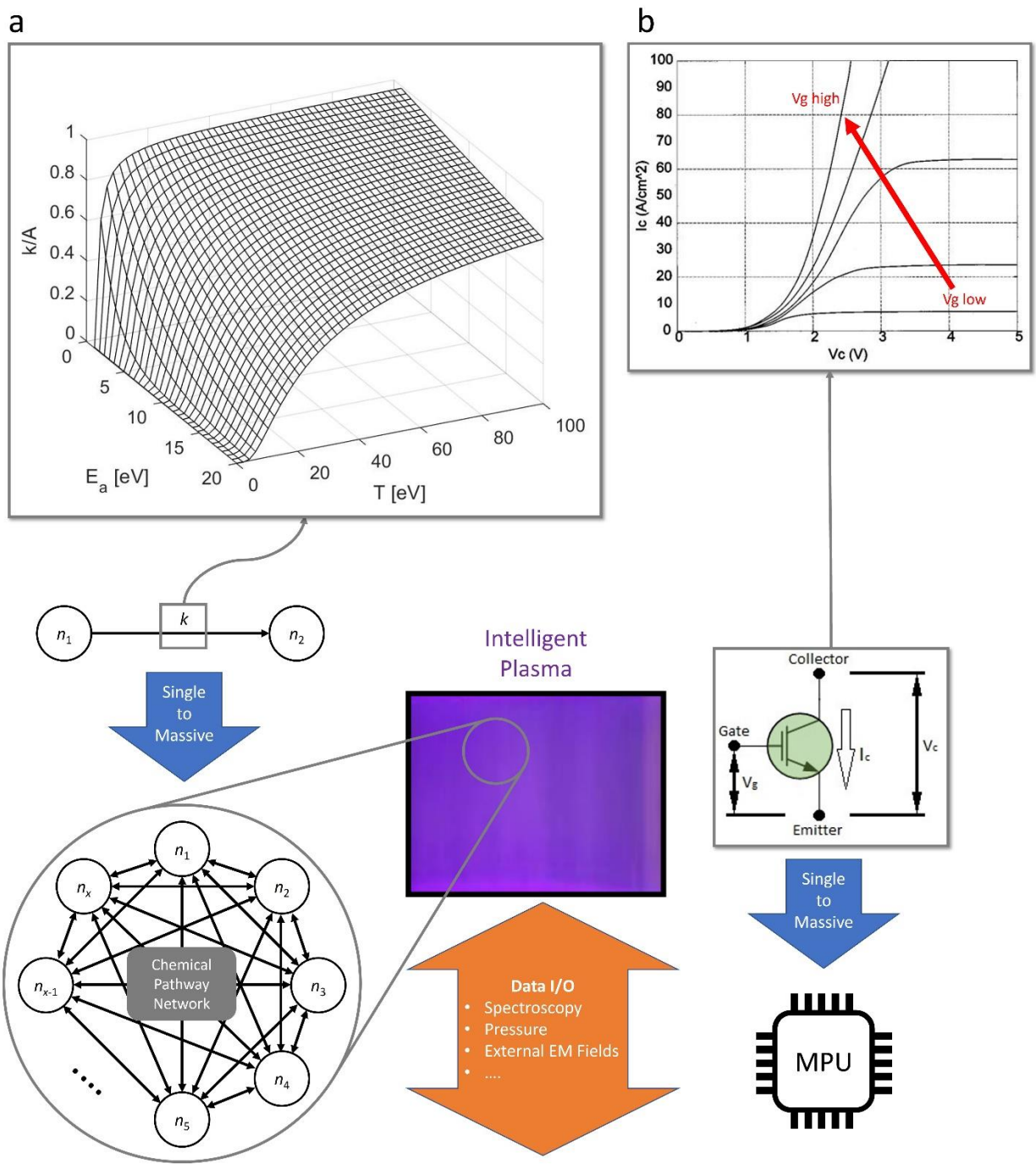

**Fig. 1 | From a single computational unit to a computational network. a**, A single chemical reaction pathway with an Arrhenius rate coefficient as an activation, and a chemical pathway network including massive chemical reaction pathways. **b**, A

single switch transistor with gate voltage controlling the collector-emitter current, and a micro processing unit including massive semiconductor components. Reproduce with permission from K. Sheng *et al.*, *IEEE T. Power Electro.* **15**, 1250-1266 (2000)[22].

**Media of Intelligence.** The programming procedure is to customize a CPN. In a chemical system, any of the molecules have non-zero probabilities to inelastically collide with all others. This is equivalent to a network of transistors all connected but may through resistors corresponding to those relatively low collision probabilities. Therefore, the customization includes two aspects. First, removing certain precursor species. Second, manipulate the temperatures in a thermal nonequilibrium system and collision frequencies. As programmable intelligent material, the data output relies on diagnostics technologies, such as the spectroscopy of photon emissions, the measurement of species temperatures, and species concentrations. The data input and the programming procedure can be the manipulations of chemical compositions and energy injections such as pressure change and electromagnetic (EM) waves. Therefore, a generally appropriate media to be programmed should be fluid for easy injection and exhaust, a matter sensitive to external EM fields, but not a condensed matter where pressure cannot be manipulated to control the molecular collision frequencies. The choice is thus plasmas which obey macroscopic magnetohydrodynamics with microscopic particle collisions and chemical reactions. Plasmas can also provide spontaneous emissions for data output.

Another attractive method is to mix supramolecules into plasmas as aerosol addons. Many examples of self-propelled nanomachines and shuttles include molecular-size cargo transportations[23-25]. The most classic and famous example is the rotaxane ring moving along an alkyl chain[27,28]. Such a mechanism can be a valve of molecular containers[29]. The container can be a more effective chemical species storage. Also, the supramolecular tweezers can be designed to capture certain chemical species[26]. The tweezers can thus be mixed in gases and plasmas to capture, save, and release species to customize CPN.

**Examples**

**Automatic Workflow.** Plasma etching is a middle procedure of modern semiconductor chip fabrications, while periodic etching is a common method to prevent sidewall damage[30]. After each etching phase, a protective $C_4F_8$ layer is applied on the silicon surface to prevent over-etching to the side during the next etching period[30,31]. To that end, an intelligent plasma combining with supramolecular containers can complete such a mission without any external controls. As shown in Fig. 2a, During the etching, the more film removed, the more etching products are released into the plasma plume and be excited emitting in a specific wavelength of photons. The cyclobis(paraquat-p-phenylene) ($CBPQT^{4+}$) as the molecular shuttle will move from the dinitrophenol (DNP) position to the tetrathiafulvalene (TTF) position and open the nano valve to release proper numbers of $C_4F_8$ molecules stored in the silica container at the bottom of the nano valve[29]. The $C_4F_8$ molecules will form a surface layer and stop the etching. Without the exciting etching product emitting photons, the release of $C_4F_8$ is thus stopped. Next, the etching will thus resume until the $C_4F_8$ layer is consumed. Therefore, the chemical system can determine by itself that if the $C_4F_8$ molecules should be released depending on the number of etching products. The flowchart of this preprogrammed chemical algorithm is summarized in Fig. 2b. The CPN is shown in Fig. 2c where $n_{ion}$ is the density of ions approaching the Si substrate and mask, $n_s$ is the density of substrate molecule involved in the ion bombardment, $n_p$ is the ion-bombardment product, $n_{ex}$ is the one of exciting product, $n_{hv}$ is the density of photon emitted, $n_{C4F8}$ is the density of $C_4F_8$, $n_{op}$ represents the density of other product from the ion-$C_4F_8$ collisions, $n_{DNP}$ and $n_{TTF}$ are the densities of supramolecular valves at the DNP and TTF states, respectively. The CPN indicates that the ion collides with the substrate, $C_4F_8$, and the bombardment product. The more $C_4F_8$ released, the higher ion-substrate collision probability is occupied, that the mechanism of an "if" condition. Considering the mathematics of CPN in this case, Eq. (1) yields

$$\frac{\partial n_P}{\partial t} = k_1 n_{ion} n_s - k_2 n_{ion} n_P + k_3 n_{ex} \qquad (5a)$$

$$\frac{\partial n_{C4F8}}{\partial t} = k_4 n_{DNP} n_{hv} - k_5 n_{ion} n_{C4F8} \tag{5b}$$

Also, the ratio of the photon absorption rate over its generation rate

$$R_p = \frac{k_4 n_{DNP} n_{hv}}{k_3 n_{ex}} < 1 \tag{6}$$

Combining Eq. (5a), Eq. (5b), and Eq. (6), yields

$$\frac{\partial n_{C4F8}}{\partial t} = R_p \frac{\partial n_P}{\partial t} + R_p k_2 n_{ion} n_P - R_p k_1 n_{ion} n_S - k_5 n_{ion} n_{C4F8} \tag{7}$$

Considering the consumption rate of plasma ion

$$\frac{\partial n_{ion}}{\partial t} = -k_1 n_{ion} n_S - k_2 n_{ion} n_P - k_5 n_{ion} n_{C4F8} \tag{8}$$

In Eq. (7), the $C_4F_8$ concentration is proportional to one of the etching products, representing the forward propagation in the CPN. In Eq. (8), the density of the etching product is inversely proportional to $C_4F_8$, representing the layer protection in the CPN. Combining these two equations will lead to

$$\Phi(t) = \Psi(t) \frac{\partial n_P}{\partial t} + \Omega(t) n_P \tag{9a}$$

where

$$\Phi(t) = -\frac{\partial}{\partial t}\left(\frac{1}{k_5 n_{ion}} \frac{\partial n_{ion}}{\partial t} + \frac{k_1}{k_5} n_S\right) - \frac{\partial n_{ion}}{\partial t} - (R_p - 1) k_1 n_{ion} n_S \tag{9b}$$

$$\Psi(t) = \left(R_p - \frac{k_2}{k_5}\right) \tag{9c}$$

$$\Omega(t) = (R_p - 1)k_2 n_{ion} \tag{9d}$$

The solution of Eq. (9a) for $\frac{\partial n_P}{\partial t}$ is thus

$$\frac{\partial n_P}{\partial t} = \frac{-\Phi\Omega}{\Psi^2} \tag{10}$$

where the rate of etching product is proportional to $k_2$ and is inversely proportional to $k_5$. According to Eq. (7), $\frac{\partial n_{C_4F_8}}{\partial t}$ can be either positive or negative. The turning point is when the production rate of $C_4F_8$ is balanced by its consumption rate. Therefore, $\frac{\partial n_{C_4F_8}}{\partial t}$ oscillates around zero. Similar oscillation of $\frac{\partial n_P}{\partial t}$ is implied in Eq. (10), where the denominator and the term $-\Omega$ are both positive to leave the term $\Phi$ determining the sign of $\frac{\partial n_P}{\partial t}$. In Eq. (9b), the sign of $\Phi$ depends on the $C_4F_8$ protection $k_5$ comparing with the etching $k_1$. These analytical expressions of CPN thus show the programmed automatic plasma etching following the flowchart shown in Fig. 2b, while the oscillation relations derived above represent the "if" conditions.

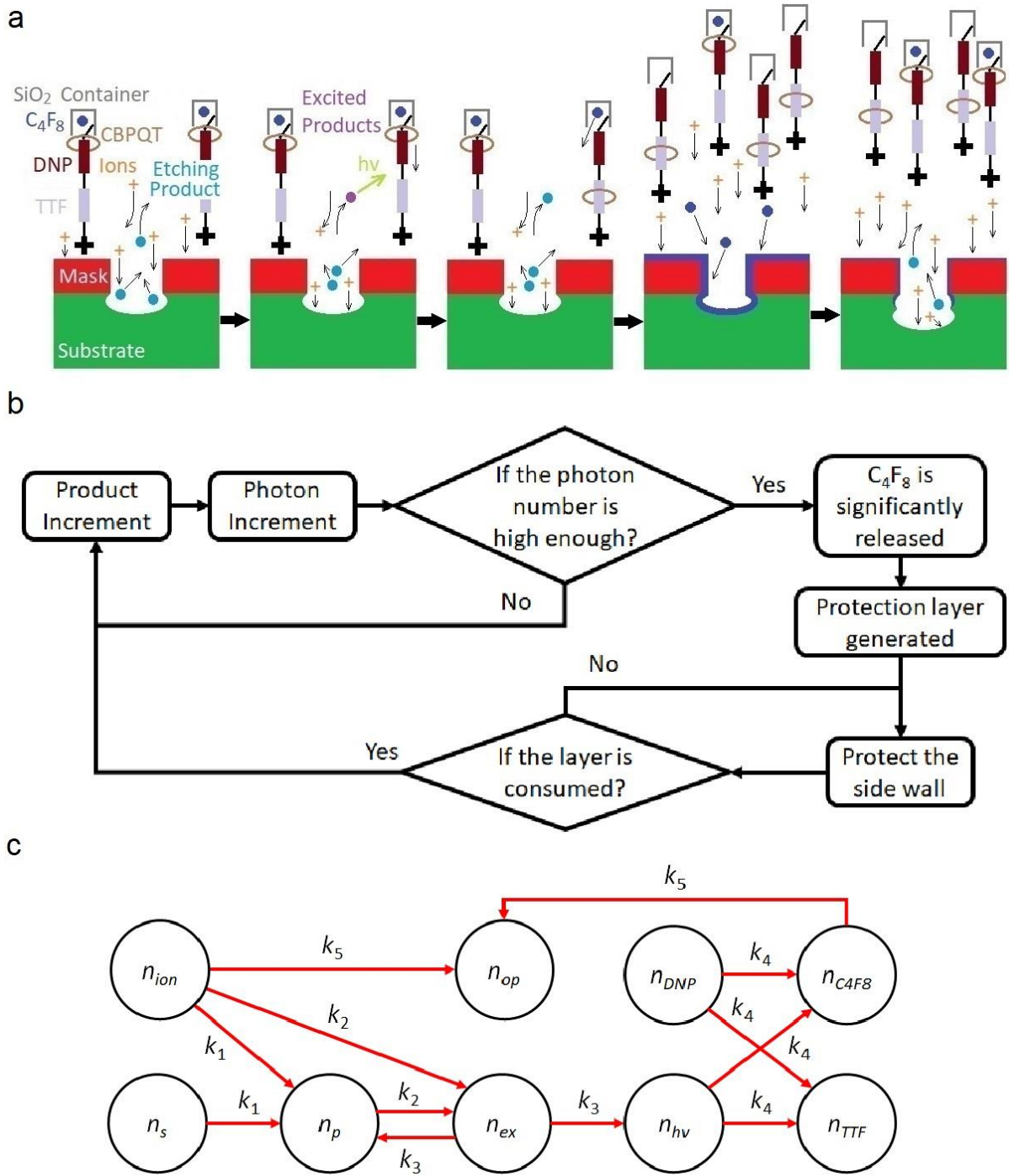

**Fig. 2 | The cyclic plasma atomic layer etching. a**, The conventional procedure where each plasma etching period will follow by a $C_4F_8$ deposition to protect the sidewall for the next etching period. **b**, The flow chart of the preprogrammed

algorithm, including two "if" conditions of etching plasma mixing with [2]rotaxane-valve molecules carrying $C_4F_8$. The valve as a molecular shuttle can be triggered by the photon emissions from the excited products. **c**, The chemical pathway network corresponding to the flow chart.

**Signal and Data Processing.** Negative permittivity can be achieved by using the split-ring discharges as 2D materials, namely "plasma-based metamaterials"[15]. One of the most fascinating applications of metamaterials is to cloak an object by twisting the optical beams[32]. Taking the EM emission in optically thin plasma as an example, the plasma frequency $\omega$ is a function of electron density $n_e$, and the permittivity $\varepsilon$[33,34]. This is an example of intelligent plasmas that automatically react to the incoming EM waves, including a Fourier transform at the molecular level. Considering a supramolecule tweezer connected to an end of a single-wall carbon nanotube (SWCNT) as shown in Fig. 3[26,35]. Each CNT-tweezer holds a guest molecule, such as benzene which has a large electron-impact ionization cross-section area. The CNT-tweezers with a variety of CNT lengths are added into the plasma. Apparently, since these supramolecules are asymmetric, they are polar molecules while additional substituents can be considered for further manipulating their polarities. When an external EM wave is applied, some of these CNT-tweezers will receive significant rotational energy. Their guest molecules can thus escape from the π-π bonds. To detailly design the CNT-tweezers, one needs molecular dynamics (MD) simulations. However, in this perspective, one can roughly evaluate the mechanism beginning with:

$$L = \int E(t) \times \sum_i q_i r_i \sin[\varphi_i + \varphi(t)] \, dt \tag{11}$$

$$E = E_0 \sin(2\pi k_E c t) \tag{12}$$

$$\frac{\partial \varphi}{\partial t} \hat{z} = L \left( \sum_i m_i r_i^2 \right)^{-1} \tag{13}$$

$$F_g = m_g \frac{\partial \varphi}{\partial t} \hat{z} \times r_g \tag{14}$$

where $L$ is the angular momentum, $E$ is the electric field of the incoming EM wave with $E_0$ for its amplitude, $k_E$ as the wavenumber, $q_i$ is the $i^{th}$ charge in the CNT-tweezer, $r_i$ is the distance from the $i^{th}$ charge to the mass center, $\varphi_i$ is the angle of the line connecting the $i^{th}$ charge and the mass center with respect to the incoming EM wave, $\varphi$ is the angle the CNT-tweezer has turned, $c$ is the light speed, $z$ is the unit vector of the rotation axis, $m_i$ is the $i^{th}$ atom mass, and $F_g$ is the extra force received by the guest molecule with its mass $m_g$ at the distance $r_g$ to the mass center. Combining these equations yields

$$F_g = m_g r_g \times \left[ -E_0 \sin(2\pi k_E c t) \times \frac{\sum_i q_i r_i \sin(\theta_i)}{\sum_j m_j r_j^2} + M_0 \right] \tag{15a}$$

$$\theta_i = \varphi_i + \varphi + \varphi_0 \tag{15b}$$

where $M_0$ is the initial angular acceleration due to the thermal velocity and $\varphi_0$ is the initial angle of the CNT-tweezer with respect to incoming EM waves. In Eq. (15a), the extra force applied on the guest molecule depends on the coupling of the oscillation of the external electric field and the rotation of the CNT-tweezer. On the other hand, the minimum force $\boldsymbol{F}_{bc}$ required for a benzene molecule to escape the π-π bond is computed using the bond energy and gap distance[36]. Making $\boldsymbol{F}_{bc}$ equal to $\boldsymbol{F}_b$ in the Eq. (15a) implies that both a too long and a too short CNT will cause insufficient rotational energy to release the guest molecule since the variable $r$ appears in both the numerator and the denominator. The shortest CNT-tweezer is higher than the lower limit, and the guest molecule number $n_g$ held by each length of CNT-tweezer is a function of the length. The mathematical relation is designed by the user, which will be discussed later. Finally, when such an intelligent plasma encounters an external EM wave, the number of free guest molecules as a unique summation of all the guest molecules released can indicate the incoming EM wave frequency. This is a discrete Fourier transform achieved at the molecular level for an EM wave with a single frequency.

Next, the guest molecules will modify the plasma chemistry and the CPN is shown in Fig. 4. The rate equations for the CPN:

$$\frac{\partial n_e}{\partial t} = k_2 n_e n_g - k_3 n_e n_{i-g} + k_4 n_e n_{gas} - k_5 n_e n_{i-gas} \qquad (16)$$

$$\frac{\partial n_{i-g}}{\partial t} = -\frac{\partial n_g}{\partial t} = k_2 n_e n_g - k_3 n_e n_{i-g} \qquad (17)$$

$$\frac{\partial n_{i-gas}}{\partial t} = k_4 n_e n_{gas} - k_5 n_e n_{i-gas} \qquad (18)$$

where $n_e$ is the electron density, $n_g$ is the guest molecule density, $n_{i-g}$ is its ion density, $n_{gas}$ is the neutral gas density, $n_{i-gas}$ is its ion density, and $k_2$ to $k_5$ is the rate coefficients as shown in Fig. 3b. Combining these equations and assuming that $k_4 n_{gas} \gg k_5 n_g$ due to the low ionization degree, we will have

$$n_g = \int k_4 n_e n_{gas} dt - \int k_5 n_e^2 dt - n_e \qquad (19)$$

Also, the plasma frequency $\omega_p$ is a function of electron density

$$\omega_p = \sqrt{\frac{n_e e^2}{\varepsilon m_e}} \qquad (20)$$

where $m_e$ is the mass of an electron. Overall, Eq. (15) provides the length $2|r|$ of CNT-tweezers which will release guest molecules, a user-designed mathematical relation determines $n_g$ as a function of the length, and combining Eq.(19) with Eq. (20) finally provides the plasma frequency from $n_g$. The equation chain shows how a plasma frequency can automatically respond to an incident EM wave. Also, any other plasma parameters can be the goal based on the user's design of the CPN. For example, one can make the guest molecules alter the excitation rates of certain plasma species to produce a spontaneous emission spectrum as a function of the input signal. Please note that the signal processor requires a collisional intelligent plasma. If the user-required signal output must be prepared in a collisionless plasma, such as the plasma oscillation emission, a photon coupling could be used between the two steps shown in Fig. 4. For example, the guest molecules of CNT-tweezers alter the excitation rates in the collisional intelligent plasma, and the excited species emit a spectrum to remotely alter another collisionless plasma.

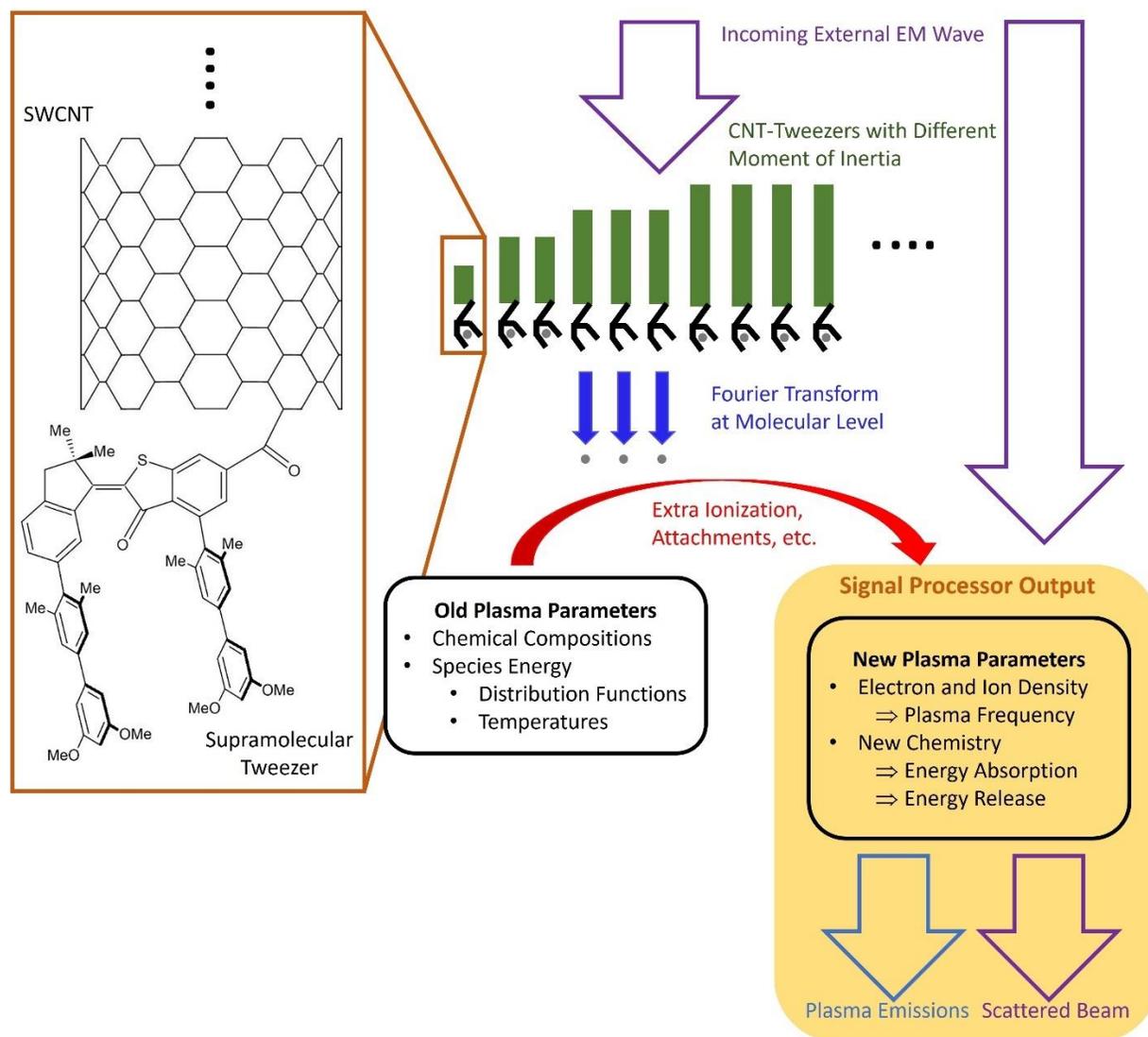

**Fig. 3 | An intelligent plasma processing an EM wave signal.** The incoming wave raises a resonance of CNT-tweezers with corresponding SWCNT length and lose their guest molecules which will further involve the plasma chemistry for signal outputs such as emissions and scattering.

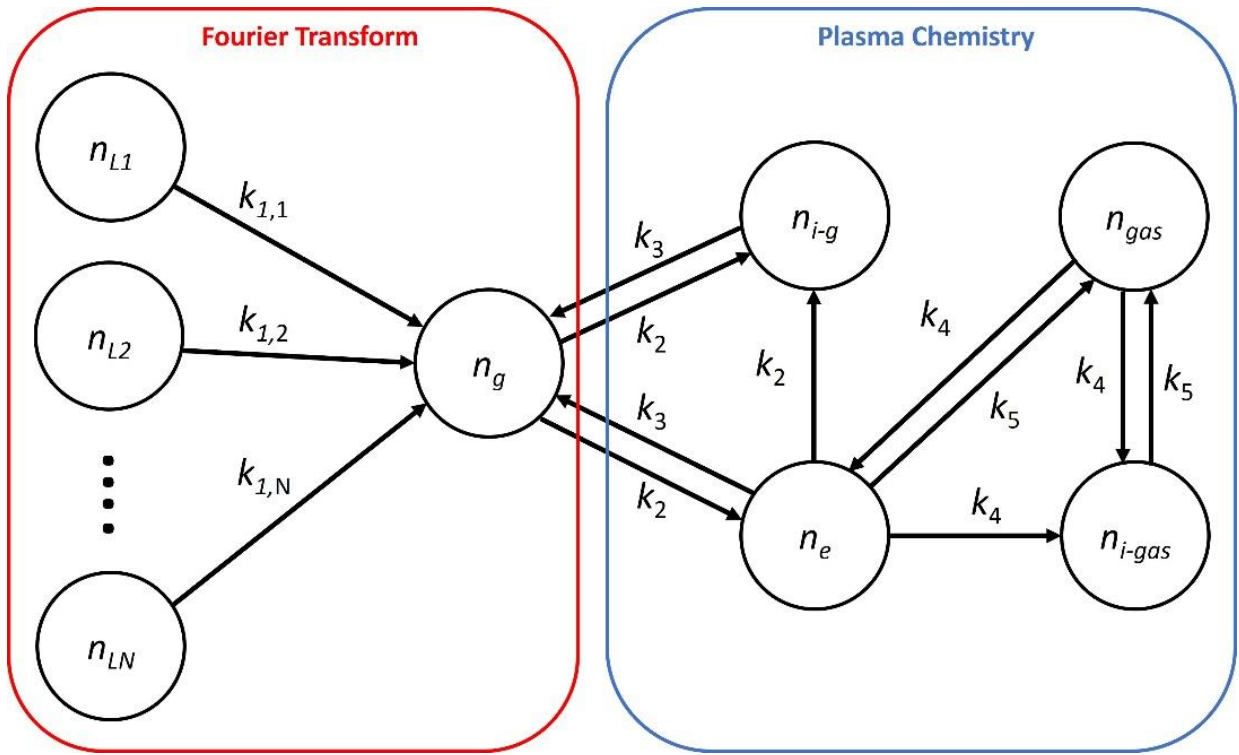

**Fig. 4 | The CPN of an intelligent plasma signal processor.** The CNT-tweezers on the left with length from $L_1$ to $L_N$ will release guest molecules to manipulate the plasma chemistry on the right. The Fourier transform stage is a signal processing unit that receives the incoming EM wave, while the plasma chemistry stage is a signal output unit to emit a new EM wave or provide certain plasma parameters representing the output signal.

## Concluding Remarks

Due to the virtually infinite ways to customize the CPN and the possibilities to program the intelligent plasma, we expect unprecedented future potentials of such media. Overall, in this perspective, we propose a practical approach to a new phase of matter with intelligence. Such novel interdisciplinary research field supports and broadens many fundamental science and industrial applications such as biomedicine, material processing, communication, and information technologies.